# Polarization-tuneable excitonic spectral features in the optoelectronic response of atomically thin ReS$_2$.

Daniel Vaquero [1,†], Olga Arroyo-Gascón [2,3,†], Juan Salvador-Sánchez [1,†], Pedro L. Alcázar-Ruano [3], Enrique Diez [1], Ana Perez-Rodríguez [1], Julián D. Correa [4], Francisco Dominguez-Adame [3], Leonor Chico [3], Jorge Quereda [3].

[1] Nanotechnology Group, USAL–Nanolab, Universidad de Salamanca, E-37008 Salamanca, Spain.

[2] Instituto de Ciencia de Materiales de Madrid, CSIC, E-28049 Madrid, Spain.

[3] GISC, Departamento de Física de Materiales, Universidad Complutense de Madrid, E-28040 Madrid, Spain.

[4] Facultad de Ciencias Básicas, Universidad de Medellín, Medellín, Colombia.

[†] These authors contributed equally to the work

**Abstract:** The low crystal symmetry of rhenium disulphide (ReS$_2$) leads to the emergence of dichroic optical and optoelectronic response, absent in other layered transition metal dichalcogenides, which could be exploited for device applications requiring polarization resolution. To date, spectroscopy studies on the optical response of ReS$_2$ have relied almost exclusively in characterization techniques involving optical detection, such as photoluminescence, absorbance, or reflectance spectroscopy. However, to realize the full potential of this material, it is necessary to develop knowledge on its optoelectronic response with spectral resolution. In this work, we study the polarization-dependent photocurrent spectra of few-layer ReS$_2$ photodetectors, both in room conditions and at cryogenic temperature. Our spectral measurements reveal two main exciton lines at energies matching those reported for optical spectroscopy measurements, as well as their excited states. Moreover, we also observe an additional exciton-like spectral feature with a photoresponse intensity comparable to the two main exciton lines. We attribute this feature, not observed in earlier photoluminescence measurements, to a non-radiative exciton transition. The intensities of the three main exciton features, as well as their excited states, modulate with linear polarization of light, each one acquiring maximal strength at a different polarization angle. We have performed first-principles exciton calculations employing the Bethe-Salpeter formalism, which corroborate our experimental findings. Our results bring new perspectives for the development of ReS$_2$-based nanodevices.

## 1. Introduction

Layered transition metal dichalcogenides (TMDs) have attracted enormous attention in the last decade due to their great potential for optics and optoelectronics.[1,2] The effect of quantum confinement in these materials, combined with a reduced electrostatic screening, results in long-lived excitonic states,[3] which dominate their optical response. Further, TMDs also present valley-dependent optical selection rules, which allow to selectively generate excitons in a given valley simply by tuning the polarization of incident light.[4,5]

To date, the most studied TMDs are those containing a group VI transition metal, i.e., $MoS_2$, $MoSe_2$, $WS_2$ and $WSe_2$, mainly because they present excitonic transitions with remarkably strong oscillator strengths and narrow bandwidths. However, in the last years, Re-based TMDs ($ReS_2$ and $ReSe_2$) have gained increasing attention. These materials also present strong exciton transitions and, differently from group VI crystals, their reduced crystal symmetry leads to optical anisotropy.[6-8] Excitonic features in the photoluminescence and differential reflectance spectra of $ReS_2$ present linear dichroism, with different exciton transitions acquiring their maximal intensity for different polarizations of light.

While the band structure and optical spectrum of atomically thin $ReS_2$ have been studied by several research groups, there are still many open questions regarding its fundamental properties. For example, different works disagree on the nature of the fundamental bandgap (either direct or indirect), as well as in the labelling of the different excitonic transitions. Discussion has also arisen regarding the actual crystalline structure of $ReS_2$, with two different growth directions reported experimentally,[9,10] both with very similar total energies.[11] Additionally, $ReS_2$ multilayers have been lately found to present two stable stacking orders at room temperature,[12,13] with substantially different optical properties. Thus, the apparently contradictory results reported in the literature for the bandgap and optical response of $ReS_2$ crystals may be related to the existence of multiple stable crystalline structures, each with different band dispersions. The optoelectronic properties of $ReS_2$ are also not fully characterized yet. In particular, current literature lacks detailed information on the spectral dependence of photoresponse in $ReS_2$-based devices, of crucial importance for several technological applications.

In this work we focus on the optoelectronic response of few-layer $ReS_2$ phototransistors. The strong Coulomb interaction and the low symmetry of the crystalline structure of $ReS_2$ have a significant impact on the optical response of the photodetectors, which is dominated by exciton transitions even at moderate temperature. By resorting to low-temperature photocurrent spectroscopy,[14,15] we identify three main excitonic features, whose intensities modulate as a function of the polarization of incident light. To uncover the origin of the excitonic features observed in the spectra, we have also performed density functional theory (DFT) first-principles calculations to obtain the band structure of ReS$_2$. From this band structure, exciton absorption spectra have been obtained, allowing us to identify the excitonic origin of the various features observed in the optical spectra.

## 2. Results

### 2.1. Device and measurement geometry

The crystal structure of monolayer ReS$_2$ presents a distorted 1T structure [16], where two nonequivalent perpendicular directions are present, either parallel or perpendicular to the *b* crystalline axis. In multilayer crystals, the stacking between layers presents two different stable configurations, named in the literature as AA and AB stacking.[12,13] For AA-stacked crystals the successive layers are positioned directly on top of each other, while for AB stacking consecutive layers are shifted by roughly 2.5 Å across the *a* crystalline axis. The orientation of the ReS$_2$ crystalline axes, as well as the stacking configuration can be revealed by polarization-resolved Raman spectroscopy (see Supplementary Note 1) and photoluminescence spectroscopy (Supplementary Note 2).

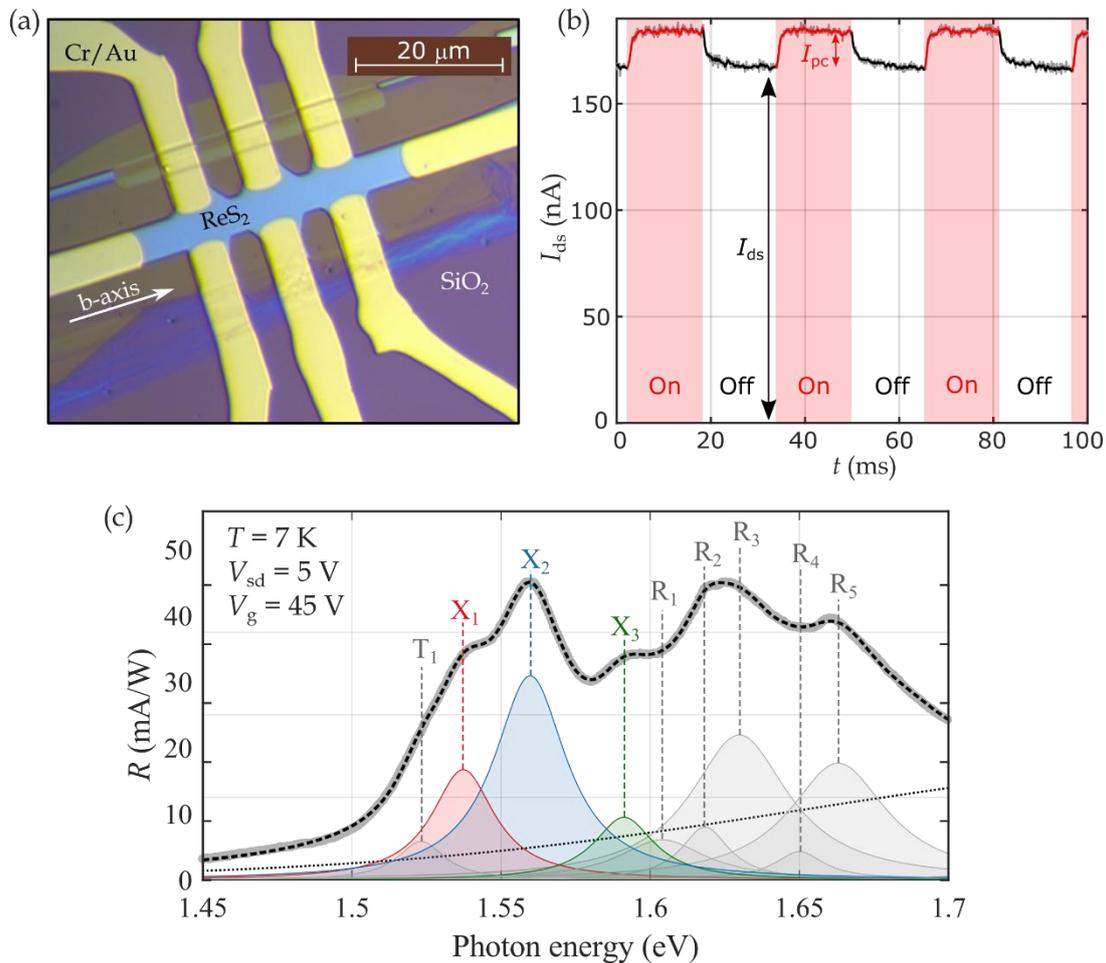

**Figure 1.** Optoelectronic response of the device. (a) Optical microscopy image of the ReS$_2$ photodetector (b) Source-drain current of the photodetector at $V_{sd}$ = 5 V and $V_g$ = 45 V. When the light excitation ($\lambda$ = 620 nm) is turned on, the drain-source current increases by $I_{PC}$. (c) Photocurrent spectrum at $V_{sd}$ = 5 V, $V_g$ = 45 V and a power density of 500 W m$^{-2}$. The spectrum is acquired for light polarization perpendicular to the *b*-axis of the ReS$_2$ crystal. The grey curve is the experimentally measured spectrum. The dashed black curve is a least square fit to a multi-Lorentzian function plus a smooth background, which accounts for direct band-to-band absorption. The individual Lorentzian peaks are also shown in the figure.

In order to explore the optoelectronic properties of few-layer ReS$_2$, we fabricate field-effect phototransistors using mechanically exfoliated few-layer ReS$_2$ crystals as the semiconductor channel. Figure 1a shows a microscope image of a typical ReS$_2$ device, fabricated with an AB-stacked ReS$_2$ crystal. Detailed discussion on the device fabrication and electrical characterization are provided Supplementary Notes 3 and 4, respectively.

To characterize the photoresponse of ReS$_2$ we expose the whole area of the device to homogeneous monochromatic light and measure the drain-source current at fixed gate and drain-source voltages. We define the photocurrent $I_{PC}$ as the difference between the drain-source current measured under illumination and in the dark (see Figure 1b).

We find that the device photoresponse is relatively fast, with response times in the order of 1 ms (see Supplementary Note 5). It is also worth remarking that we did not observe any long-lasting photodoping effects, even when they are frequently found in 2D material-based devices.

*2.2.    Photocurrent spectroscopy measurements*

Next, we explore the exciton physics of few-layer ReS$_2$ by low-temperature photocurrent spectroscopy. Unless otherwise stated, all the measurements presented here are acquired at 7 K using the AB-stacked ReS$_2$ device shown in Figure 1a. Room-temperature measurements and measurements for an AA-stacked crystal are provided in Supplementary Notes 6 and 7, respectively. We use a lock-in amplifier to register the photocurrent in the device while continuously switching the illumination on and off at a fixed frequency of 31.81 Hz. The photoresponsivity spectra are obtained by repeating this measurement while scanning the illumination wavelength. Further details regarding the procedure for spectral acquisition can be found in Supplementary Note 8. Figure 1c shows a typical ReS$_2$ photoresponsivity spectrum, acquired for light polarization perpendicular to the *b* crystalline axis. The experimentally measured photoresponse presents numerous exciton-like features in the energy range from 1.45 eV to 1.7 eV, in good agreement with the spectral features reported in literature for low-temperature photoluminescence spectroscopy in ReS$_2$.[18,19]

Our spectra present two prominent peaks, X$_1$ and X$_2$, at roughly 1.54 and 1.56 eV (highlighted in Figure 1c in red and blue, respectively). These two spectral features have also been observed in earlier literature by low-temperature PL, PLE and micro-reflectance spectroscopy.[7,18,20] Ab-initio calculations have shown that ReS$_2$ has two direct bandgap minima with very similar energies, occurring at the K$_1$ and Z points of the reciprocal lattice.[21] Thus, X$_1$ and X$_2$ are usually attributed to exciton absorption at K$_1$ and Z, respectively.[22] In our spectra, we also observe a smaller satellite peak at 1.525 eV, 14 meV below X$_1$, which we attribute to the trion state T$_1$ associated with X$_1$. A third, less prominent peak is also observed at 1.588 eV. We believe that this feature, not reported in earlier literature, may be related with a higher-energy exciton transition, which we label as X$_3$ (further discussed below). Finally, a cluster of exciton-like peaks appears at energies between 1.6 and 1.7 eV, which we attribute to the Rydberg series of excited states of the main excitons. These secondary peaks, as well as the main excitonic features, are also present in our DFT simulations of the absorption spectrum of ReS$_2$. The energies of the excitonic peaks obtained theoretically are in good agreement with our experimental findings, as explained in the Discussion section.

The dashed line in Figure 3 shows a fitting of the experimental spectrum to a multi-Lorentzian function. The function includes individual Lorentzian peaks to account for the T$_1$, X$_1$, X$_2$ and X$_3$ peaks, as well as 5 additional peaks (labeled here as R$_1$-R$_5$) to account for the most prominent spectral features between 1.6 and 1.7 eV. It must be noted that, since peaks R$_1$ to R$_5$ are not fully resolved in our spectra, it is possible that a larger number of optical transitions are present at this energy range. Thus, more than one optical process may be contributing to photoresponse for each of the assigned peaks. In addition to the different Lorentzian peaks, our fitting function also includes a Fermi-Dirac distribution function centered at 1.7 eV (shown as a dotted line in the figure) to account for the effect of direct interband transitions.

### 2.3. *Polarization dependence of photocurrent spectra*

In order to achieve a clearer picture of the exciton physics in ReS$_2$, we next characterize the dependence of photocurrent spectra on the polarization of the optical excitation. Figure 2a shows a series of photocurrent spectra acquired for different angles of polarization. All the spectral features described above are strongly affected by the polarization direction. Figure 2b shows the intensity of the three main exciton lines (X$_{1-3}$) as a function of the direction of polarization. Exciton X$_1$ becomes maximal when light polarization is at 10° relative to the *b* axis of the crystal, while X$_2$ maximizes at 95° polarization, almost perpendicular to X$_1$. The spectral feature at 1.525 eV follows the same polarization dependence of X$_1$, supporting its labelling as the trion state T$_1$. Finally, exciton X$_3$ becomes maximal for a polarization angle of 65°. The fact that the polarization dependence of X$_3$ is different from those of X$_1$ and X$_2$ supports the idea that X$_3$ does not correspond to an excited state of neither X$_1$ nor X$_2$, but it is indeed originated from an exciton transition at a different point of the reciprocal lattice. As shown in panels *c* to *e* of Figure 2, the remaining spectral features with energies above 1.6 eV follow the same polarization dependence as one of the three main exciton peaks, suggesting that they are originated by excited states of either of the main excitons. In particular, we find that $R_1$ emulates the same polarization dependence of $X_1$, while $R_2$, $R_3$ and $R_5$ emulate the dependence of $X_2$. $R_4$ becomes maximal at an angle of 40°, and probably contains contributions of excited states from both $X_1$ and $X_3$.

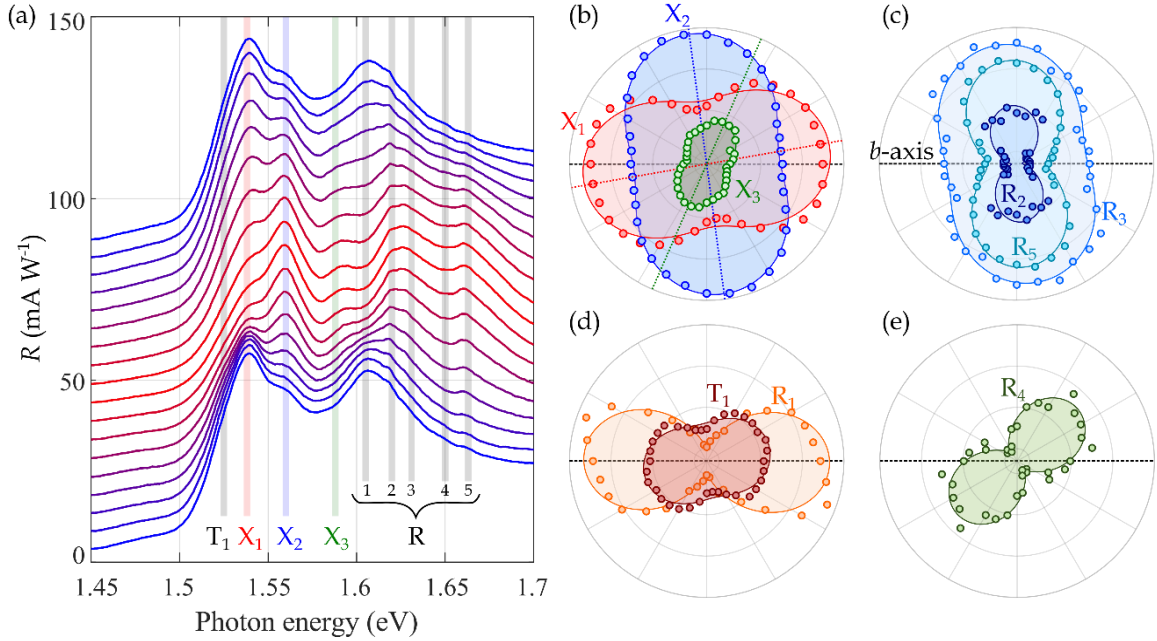

**Figure 2.** (a) Photocurrent spectra acquired for different angles of light polarization, from 0 to 170° relative to the $b$ crystalline axis, with $V_{sd}$ = 5 V, $V_g$ = 45 V and a power density of 500 W m$^{-2}$. Consecutive spectra have been shifted vertically in steps of 5 mA W$^{-1}$ for easier visualization. (b-e) Polar plots showing the modulation of the different spectral features as a function of the polarization direction, extracted from fittings to multi-Lorentzian curves, as the one shown in Figure 1c.

Our results are highly compatible with earlier reports on low-temperature photoluminescence spectroscopy for few-layer ReS$_2$. Indeed, said reports also revealed the $X_1$ and $X_2$ spectral features at very similar energies to the ones obtained here, as well as higher-energy features corresponding to excited states of these two excitons. However, the spectral feature labelled here as $X_3$ is not observed in earlier reported photoluminescence spectra. Thus, we believe that this feature may correspond to a non-radiative exciton level (further discussed below).

*2.4.    Role of in-plane electric field on the photocurrent spectra*

We now turn our attention to the effect of the drain-source voltage in the photocurrent spectra. Since excitons are charge-neutral, they need to dissociate into free electron-hole pairs in order to contribute to photoresponse. In typical 2D phototransistors, this exciton dissociation process mainly takes place in the vicinity of the electrodes, where the presence of Schottky barriers results in very strong in-plane electric fields.[23] Figure 3 shows five individual photocurrent spectra, acquired at different source-drain voltages, ranging from 1 V to 5 V. To facilitate comparison, the five spectra have been normalized to the value of the photocurrent at 1.9 eV, where photoresponse should be mainly caused by direct interband absorption. As expected, when we increase the source-drain voltage (and therefore the in-plane electric field) exciton dissociation processes become stronger, and the excitonic spectral features become progressively more prominent relative to the smooth background caused by interband transitions. This effect is particularly noticeable for the $X_2$ peak, which is barely visible at low $V_{sd}$ due to the presence of more prominent spectral features in its vicinity, but becomes clearly resolved at $V_{sd}$ = 5 V. Furthermore, we observe that the relative intensity of the excitonic features is also affected by $V_{sd}$, with the $X_{1-3}$ peaks becoming more prominent as $V_{sd}$ increases, relative to the $R_{1-5}$ peaks. This was also expected, as the excited Rydberg states have weaker binding energies compared to the fundamental states and, therefore, exciton dissociation for these states may occur at weaker electric fields.

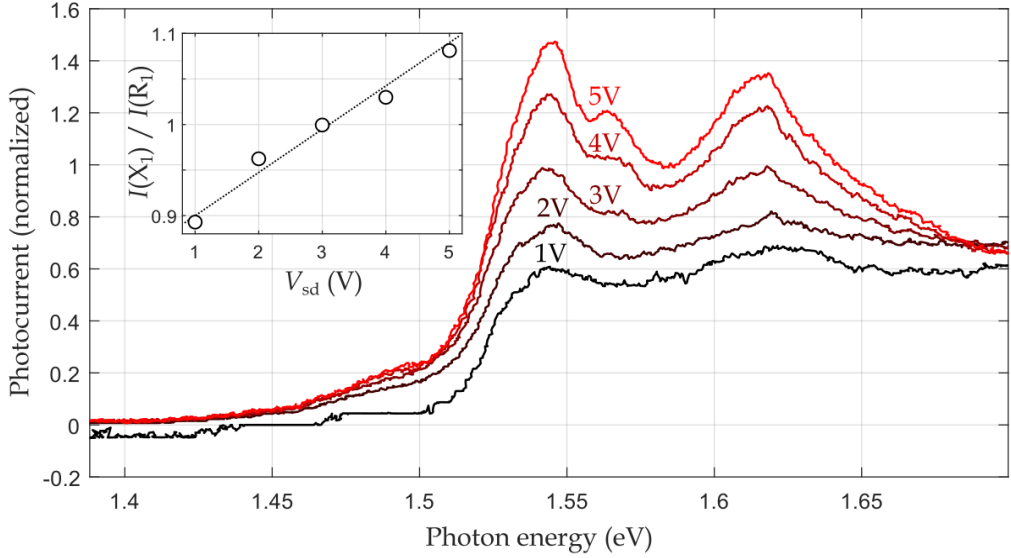

**Figure 3.** Photocurrent spectra acquired for different $V_{sd}$, ranging between 1 V and 5 V, for $V_g$ = 45 V, a power density of 500 W m$^{-2}$, and light polarization parallel to the b-axis of the ReS$_2$ crystal. To facilitate comparison between the different spectra, they have been normalized to the value of photoresponse at 1.9 eV. The inset shows the ratio between the photocurrent at 1.54 and 1.62 eV, on resonance with peaks $X_1$ and $R_1$, respectively. As $V_{sd}$ is increased, this ratio becomes progressively larger.

### 2.5. Gate voltage dependence of the photocurrent spectra

Finally, we characterize the gate voltage dependence of the photocurrent spectra. Figure 4a shows a set of photocurrent spectra acquired at different gate voltages ($V_g$), ranging from 24 to 50 V, for polarization parallel to the b-axis of the ReS$_2$ crystal. As expected for photogating, the overall photoresponse of the device decreases as $V_g$ is lowered. Thus, to facilitate comparison of spectral features, each spectrum has been normalized to the value of photocurrent at 1.8 eV, far away from the exciton transitions. Three main effects can be observed here: Firstly, the overall intensity of excitonic peaks relative to the smooth background increases with the gate voltage, suggesting that exciton dissociation processes become more efficient at higher gate voltages, possibly due to an increased scattering with conduction-band electrons.

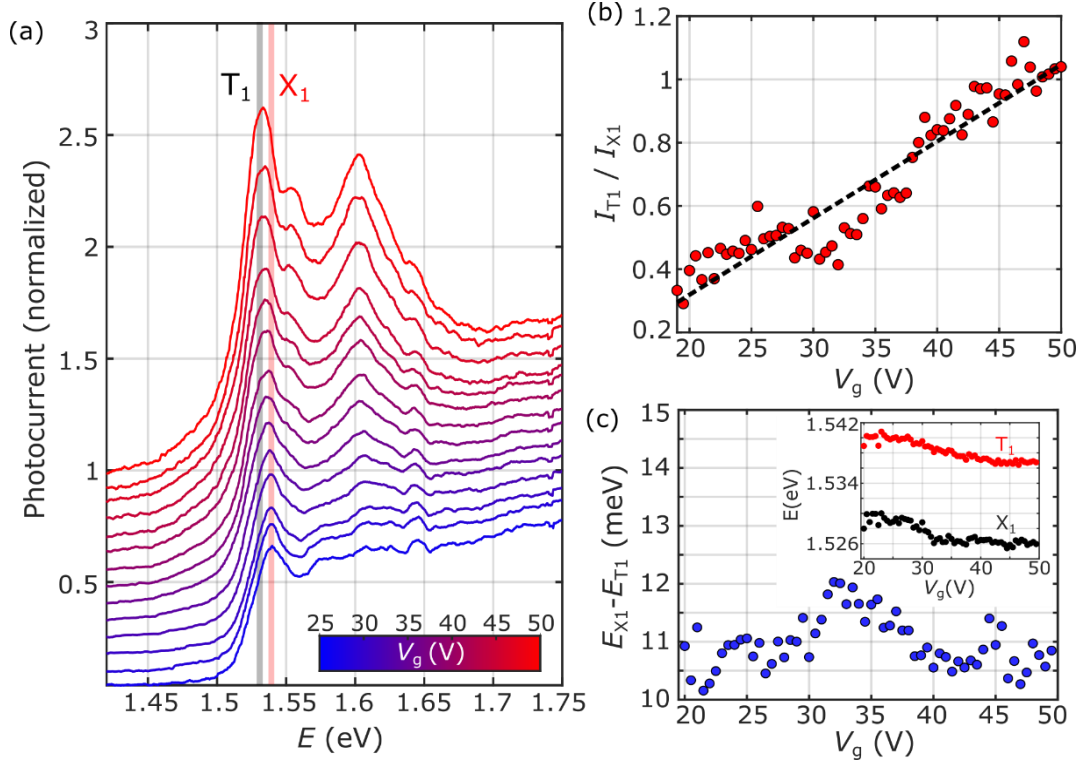

**Figure 4.** Gate dependence of photocurrent spectra. (a) Photocurrent spectra acquired for different $V_g$, ranging between 24 V and 50 V, for $V_{sd}$ = 5 V, a power density of 500 W m$^{-2}$, and light polarization parallel to the b-axis of the ReS$_2$ crystal. To facilitate comparison between the different spectra, they have been shifted vertically relative to each other, and the intensity of each individual spectrum has been normalized to the value of photoresponse at 1.8 eV. (b) Ratio between the intensities of the $X_1$ and $T_1$ spectral peaks as a function of $V_g$. (c) Energy difference between the $X_1$ and $T_1$ spectral peaks, extracted from multi-lorentzian fittings of the spectral profiles, as a function of $V_g$. The inset shows the individual energy values obtained for $X_1$ and $T_1$.

Secondly, the relative weight of the higher-energy peaks labelled as $R_1$-$R_5$ modulates in a highly nontrivial manner. While we do not fully understand the origin of this complex gate modulation of Rydberg states, a similar effect has been also observed in the photocurrent spectra of monolayer MoS$_2$.[14,15]

Lastly, we find that the relative intensity of the $X_1$ and $T_1$ peaks is also affected by the gate voltage. Figure 4b shows the ratio between the intensities of $T_1$ and $X_1$ as a function of $V_g$. When the $V_g$ is increased, the spectral weight is transferred from the neutral exciton state to the trion state, and the ratio $I_{T1}/I_{X1}$ increases monotonically. This modulation suggests that $T_1$ corresponds to a negatively charged trion state.

3. Discussion

As discussed in the introduction, both experimental and theoretical works on ReS$_2$ span a variety of results, sometimes contradictory. The first discrepancy arises from the report of two possible triclinic structures, either grown along the *c*-axis,[9] or along the *a*-axis.[10] These two different structures present slight (but not negligible) differences as to their bands and the direct or indirect character of their gaps, which have called the attention of several research groups.[11,24-26] But since the difference in total energy is very small,[11] these results are compatible with the fact that both structures can be grown depending on the substrate and experimental conditions. In any case, the theoretically reported bulk gaps range from 1.32 eV by Gadde et al.[24] to 1.6 eV by Echeverry et al.[27] for the *c*-axis and *a*-axis structure respectively, but at different points of the Brillouin zone. In fact, Gadde et al. have performed a numerical comparison between the two crystalline structures and report an indirect gap for the a-axis structure of 1.49 eV.[24] Interestingly, other authors have pointed out that a smaller gap, from 1.2 eV to 1.29 eV, can be obtained if the whole band structure is studied, and not only along the high-symmetry paths and edges of the Brillouin zone, as it is customarily done.[25] This observation is relevant given the low symmetry of this material, and it has also been reported in other low-symmetry systems.[28] Nevertheless, since many-body effects are neglected in most calculations, it is expected that not only the gap character may change from direct to indirect if these effects are included, but also an increase of its value and a flattening of the bands.

Table I. Exciton absorption energies of the experimental data fit (column 2) and DFT results for several functionals (columns 3 to 6). The last two columns gather other theoretical works, where only the first peaks were discussed. The feature at 1.40 eV is described in the literature as an indirect transition.[13,27] All units are in eV.

| Label | Exp. | PBE | vdW-DF | vdW-DF2 | TB09 | Ref. 13 | Ref. 27 |
|---|---|---|---|---|---|---|---|
| – | – | – | – | – | – | 1.40 | 1.40 |
| – | – | – | – | – | – | 1.46 | – |
| $T_1$ | 1.524 | – | 1.514 | 1.50 | 1.471 | 1.50 | 1.51 |
| $X_1$ | 1.537 | 1.538 | 1.543 | 1.535 | 1.535 | – | – |
| $X_2$ | 1.558 | 1.562 | – | – | 1.575 | 1.56 | 1.56 |
| $X_3$ | 1.590 | – | 1.602 | 1.590 | 1.592 | – | – |
| $R_1$ | 1.607 | – | – | 1.604 | 1.60 | – | – |
| $R_2$ | 1.619 | 1.616 | 1.621 | 1.614 | 1.621 | – | – |
| $R_3$ | 1.631 | 1.629 | 1.628 | 1.633 | 1.636 | – | – |
| $R_4$ | 1.649 | 1.645 | – | 1.644 | 1.641 | – | – |
| $R_5$ | 1.660 | 1.654 | 1.662 | 1.656 | 1.665 | – | – |

Since our experimental samples follow the *a*-aligned crystal structure, we have performed our DFT simulations employing such geometry. Several exchange-correlation functionals and methods were tested, yielding indirect gaps between 1.24 eV to 1.26 eV, in agreement with the former works discussed above, except for the TB09 functional which results in a 1.425 eV gap, in a similar way to other meta-GGA functionals (see Methods). In Supplementary Note 9 we present the band structure of ReS$_2$ computed with the PBE functional. A detailed comparison of the exciton absorption energies obtained via the Bethe-Salpeter equation for all tested functionals is gathered in Table I.

Due to the low symmetry of ReS$_2$, the number of atoms in our DFT unit cell makes this material a good candidate for meta-GGA calculations. Since GW calculations result in an overall uniform shift of the bands, different scissor shifts are performed for each column on Table I, so that the gap of each functional matches the GW one,[27] as detailed in the Supplementary Information. We find a general agreement between theoretical and experimental data for all functionals. The TB09 calculation reproduces successfully all the main excitonic features, including the newly reported transition at around 1.59 eV, labeled X$_3$ in the former section. However, the X$_2$ feature is best described by the PBE functional. Thus, we conclude PBE and TB09 yield the most precise results among all functionals studied, being the TB09 calculation the excelling one, with a thorough description of all experimental peaks.

In all, the spectral characterization of ReS$_2$ photoresponse reported here shines new light on the polarization-dependent optoelectronic response of excitons on ReS$_2$. In particular, the observation of a novel exciton transition in our photocurrent spectra, not reported in earlier literature for optical spectroscopy measurements opens new possibilities for studying and exploiting excitonic phenomena in ReS$_2$-based optoelectronic devices. However, further work is needed to fully clarify the origin and properties of this newly observed excitonic feature.

**Methods**

*Photocurrent spectral acquisition* – The sample is placed inside a closed cycle cryostat at T=7 K with an optical access and exposed to laser illumination. The light source is a supercontinuum (white) laser (SuperK Compact), and the excitation wavelength is selected using a monochromator (Oriel MS257 with 1200 lines/mm diffraction grid). This allows to scan the visible and NIR spectral range, roughly from 450 nm to 1000 nm. The polarization of the light is selected using a linear polarizer and a half-waveplate. Electrical measurements are performed with a doubled channel sourcemeter (Keithley 2614b). For AC optoelectronic measurements, the optical excitation is modulated by a mechanical chopper and the photoresponse of the device is registered using a lock-in amplifier (Stanford Research SR830).

*Raman and photoluminescence spectroscopy* – For Raman and photoluminescence spectroscopy measurements we use a Horiba LabRam HR micro-Raman spectrometer with a 100x objective under 532 nm laser excitation in normal incidence on the sample. Raman polarization-resolved measurements are performed rotating the sample while the excitation and detected light have the same polarization. Photoluminescence-resolved measurements are performed fixing the sample position and changing the polarization angle of the collected light using a linear polarizer.

*First principles simulations* – DFT simulations were mainly performed using the GPAW code[29-31] with a plane wave energy cutoff of 500 eV. A 10x10x10 Monkhorst-Pack grid was used for the k-space sampling, and we considered structural relaxations until forces on the atoms were below 0.001 eV/Å. A full Brillouin zone analysis and comparison with former works[32,33] yields slightly more precise band gaps for the unrelaxed geometry, which also provides a more direct comparison with our experimental samples.

Several exchange-correlation functionals were tested under the generalized gradient approximation (GGA). In particular, the PBE parametrization and van der Waals corrections under the vdWDF and vdW-DF2 approximations were used.[34] Additionally, the meta-GGA Tran-Baha modified Becke-Johnson (TB09) functional is considered in order to render a more accurate description of the band gap.[34] The gaps for the PBE and van der Waals functionals range from 1.243 to 1.263 eV, whereas the TB09 functional gives a 1.425 eV gap. Spin-orbit coupling was shown not to have an impact in the location of direct transitions since band splittings are forbidden by inversion symmetry (see Fig. S11 in Suppl. Info.). Moreover, equivalent band structure calculations were performed resorting to the SIESTA code,[35,36] yielding a good agreement between both methods.

As for the absorbance spectrum, the Bethe-Salpeter equation as implemented in GPAW was employed. A 9x9x9 Monkhorst-Pack mesh was used for these calculations. 5 valence and 5 conduction bands were considered, as well as an energy cutoff of 100 eV. For the sake of comparison, this calculation was performed for the PBE, vdW-DF, vdW-DF2 and TB09, as shown in Table I in the main text. A scissor shift is applied to each calculation in order to reproduce the results one would obtain performing BSE over a GW method. Specifically, the applied shift amounts to 0.375, 0.349, 0.339 and 0.167 eV for the PBE, vdW-DF, vdW-DF2 and TB09 functionals, respectively.

**Author Contributions**

J.Q. conceived and supervised the research. D.V.-M., J. S.-S., A.P.-R. and P.L.A.-R. were involved in the fabrication and electrical characterization of ReS$_2$ phototransistors. D.V.-M. and A.P.-R. carried out the electronic, optoelectronic, and spectral measurements and data analysis. O.A.-G., J.D.C., L.C. and F.D.-A. performed the theoretical analysis. The article was written through contribution of all the authors, coordinated by J.Q. All authors have given approval to the final version of the manuscript.

**Funding Sources**

We acknowledge financial support from the Agencia Estatal de Investigación of Spain (Grants PID2022-136285NB, PID2019-106820RB, RTI2018-097180-B-100, and PGC2018-097018-B-I00). We acknowledge financial support by Comunidad de Madrid through the (MAD2D-CM)-UCM5 project. We acknowledge financial support by Junta de Castilla y León (Grants SA256P18 and SA121P20), including funding by ERDF/FEDER. J.Q. acknowledges the financial support received from the Marie Skłodowska Curie-COFUND program under the Horizon 2020 research and innovation initiative of the European Union, within the framework of the UNA4CAREER program (grant agreement 4129252). D.V.-M. acknowledges financial support from the Spanish Ministry of Universities (Ph.D. contract FPU19/04224). O.A.-G. acknowledges the support of grant PRE2019-088874 funded by MCIN/AEI/10.13039/501100011033 and by "ESF Investing in your future". A.P-R acknowledges the financial support received from the Marie Skłodowska Curie-COFUND program under the Horizon 2020 research and innovation initiative of the European Union, within the framework of the USAL4Excellence program (grant agreement 101034371). J. S.-S. acknowledges financial support from the Consejería de Educación, Junta de Castilla y León (EDU/875/2021), and ERDF/FEDER.

**Acknowledgments**

We thank Mercedes Velázquez for her help with the photoluminescence and Raman characterization setup. We thank Yahya M. Meziani and Adrián Martín-Ramos for their assistance on the development of the photocurrent spectroscopy measurement setup. We thank Vito Clericò for his assistance on the device fabrication process. We thankfully acknowledge the computer resources at FinisTerrae III and the technical support provided by Centro de Supercomputación de Galicia, CESGA (FI-2023-1-0038).

Supplementary Information to:

Polarization-tuneable excitonic spectral features in the optoelectronic response of atomically thin ReS$_2$.

Daniel Vaquero [1,†], Olga Arroyo-Gascón [2,3,†], Juan Salvador-Sánchez [1,†], Pedro L. Alcázar-Ruano [3], Enrique Diez [1], Ana Perez-Rodríguez [1], Julián D. Correa [4], Francisco Dominguez-Adame [3], Leonor Chico [3], Jorge Quereda [3].

[1] Nanotechnology Group, USAL–Nanolab, Universidad de Salamanca, E-37008 Salamanca, Spain.

[2] Instituto de Ciencia de Materiales de Madrid, CSIC, E-28049 Madrid, Spain.

[3] GISC, Departamento de Física de Materiales, Universidad Complutense de Madrid, E-28040 Madrid, Spain.

[4] Facultad de Ciencias Básicas, Universidad de Medellín, Medellín, Colombia.

† These authors contributed equally to the work

# 1. Crystal symmetry and polarization-resolved optical spectroscopy

Figure S1a shows the crystal structure of ReS$_2$. Single layers present a distorted 1T structure [1], where two nonequivalent perpendicular directions can be observed, either parallel or perpendicular to the $b$ crystalline axis. In multilayer crystals, the stacking between layers presents two different stable configurations, named in the literature as AA and AB stacking.[2,3] For AA-stacked crystals the successive layers are positioned directly on top of each other, while for AB stacking consecutive layers are shifted by roughly 2.5 Å across the $a$ crystalline axis.

The orientation of the crystalline axes, as well as the stacking configuration can be revealed by polarization-resolved Raman spectroscopy. Figure S1c shows a set of polarization-resolved Raman spectra acquired for an AB stacked few-layer crystal. Similar spectra for an AA stacked crystal are shown in Supplementary Note 7 for comparison. The spectra are acquired by selecting a parallel configuration of the polarization angle between the incident and collected light ($z(uu)\bar{z}$ in Porto notation, where $u$ is the polarization axis, contained in the $xy$ plane and forming an angle $\theta$ relative to the $b$ axis of the crystal). The different spectra are obtained by rotating the sample around the z axis to change the angle $\theta$. In the range between 120 and 220 cm$^{-1}$, the Raman spectra of ReS$_2$ present five main A$_g$-like active modes,[4] as labelled in the figure. The crystal orientation can be identified by observing the intensity of modes I and V, which becomes maximal for $\theta = 0°$. The stacking order of the crystal can be also inferred from the spectra by measuring the difference in Raman shift between modes I and III, which is around 13 cm$^{-1}$ for AA stacking (see Supplementary Note 7) and 20 cm$^{-1}$ for AB stacking. [14]

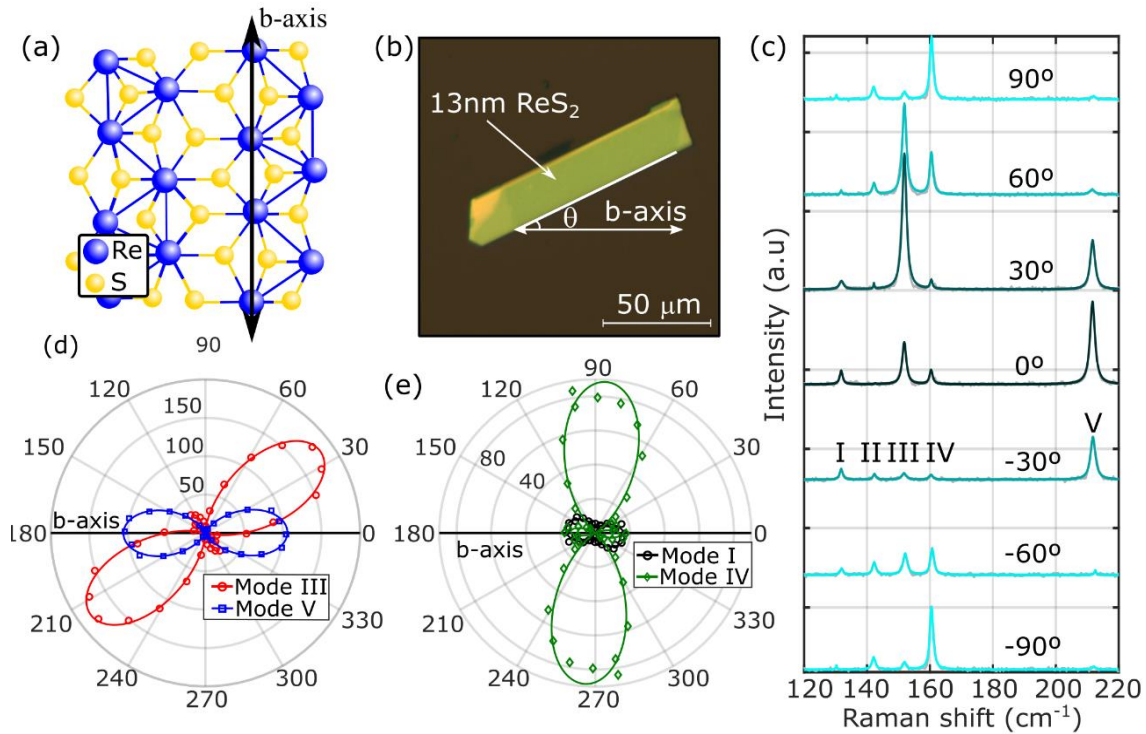

**Figure S2.** ReS$_2$ crystal symmetry and polarization-resolved Raman spectroscopy. a) Crystalline structure of ReS$_2$. b) Picture of the few-layer ReS$_2$ flake transferred onto the SiO$_2$/Si substrate. The definition of the angle between the light polarization (white arrow) and the b-axis (red line) is also shown, which will be used in the following measurements. c) Polarization-resolved Raman spectra as function of the sample orientation angle. The spectra taken every 10° between 0° and 180° are vertically offset. Mode I to V are labelled in the figure. d) Raman intensity of modes III (red circles) and V (blue squares) as a function of the polarization angle. e) Raman intensity of modes I (black circles) and IV (green diamonds) as a function of the polarization angle.

## 2. Photoluminescence spectroscopy measurements

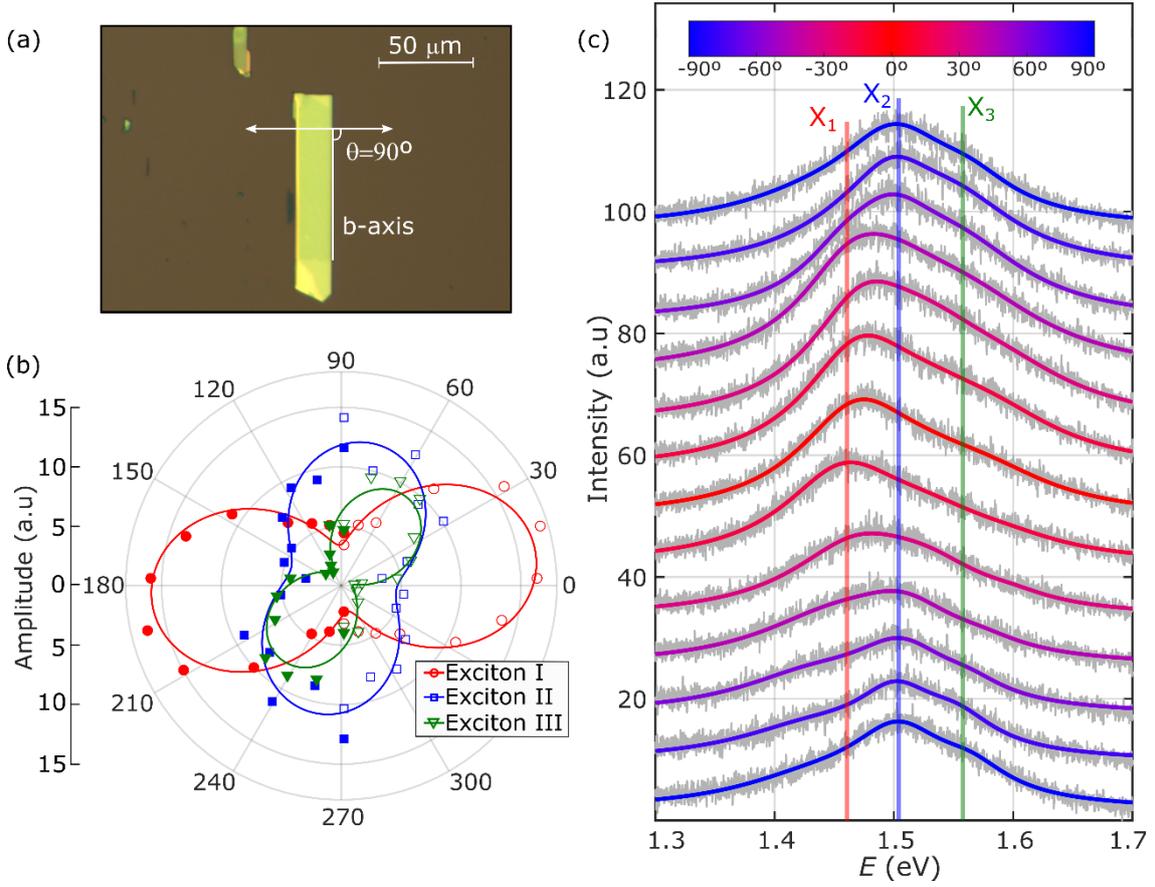

**Figure S2.** Polarization-resolved photoluminescence spectroscopy. a) Optical image of a multilayer ReS$_2$ crystal. b) Photoluminescence spectra for different output polarizations, taken every 10º between 0º and 180º. The different spectra are vertically offset in steps of 10 units to facilitate visualization. c) Intensity of the three main exciton levels X$_1$, X$_2$ and X$_3$ as a function of the polarization angle.

## 3. Main steps of device fabrication

ReS$_2$ flakes were obtained through micromechanical exfoliation of the bulk material using adhesive blue tape, followed by transfer onto a viscoelastic gel substrate based on polydimethylsiloxane (PDMS). Optical identification of the individual flakes was performed using an optical microscope, where the PDMS film was scanned by transmitted light (Figure S3a).

After selecting a suitable ReS$_2$ flake, it was transferred onto a SiO$_2$ (290nm)/Si substrate, which was cleaned using acetone and isopropanol (IPA) to remove any residual contaminants prior to the transfer. The PDMS film was deposited onto the SiO$_2$/Si substrate at room temperature and slowly removed to transfer the ReS$_2$ flake (Figure S3b). A final cleaning step was performed using acetone, IPA, and annealing with argon at 250°C for 15 min to remove any remaining contaminants from the transfer process. The thickness of the ReS$_2$ flake was 13nm, as measured using a Stylus Profilometer (Figure S3c).

The final device geometry was defined using electron beam lithography (EBL-SEM). First, homemade resist based on PMMA in chlorobenzene at 4% (by weight) was spin-coated at 4000 rpm for 1 min onto the flake and baked at 160° for 10 min. After the EBL exposure, the resist was developed with a mixture of 1 part MIBK to 3 parts of IPA. The development process was followed by a dry etching process using an ICP-RIE Plasma Pro Cobra 100 with an SF6 atmosphere (40 sccm, P=6 mTorr, P=75 W at 10°C) (Figure S3d). A second EBL process was performed using the same PMMA (4% by weight) to define the area of the planar contacts on top of the device. The contacts were deposited via evaporation of Cr/Au (5 nm/45 nm) followed by a standard lift-off procedure.

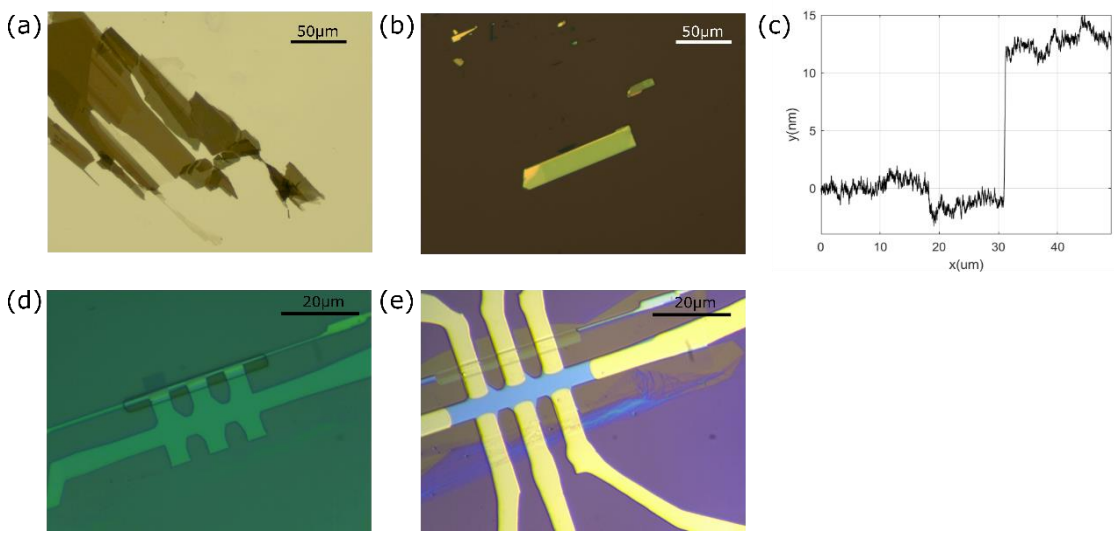

**Figure S3.** Optical images of the different steps of the device fabrication process **a.** Few-layer flake of ReS$_2$ on a PDMS film. **b.** Few-layer flake of ReS$_2$ selected for the final device transferred onto a SiO$_2$/Si substrate. **c.** Thickness profilometer measurement **d.** Final geometry of the device after the EBL and dry etching process. **e.** Final device.

The final device, forming a Hall bar with a central horizontal width of W=5 µm and a distance between contacts of L=10 µm, is shown in Figure S3e. The device was bonded on a DILL14 chip carrier for photocurrent measurements.

## 4. Electrical characterization of the device

Figure S4a shows a low-temperature gate transfer curve of the ReS$_2$ device. There, the n-type semiconductor character is clearly observed, as the semiconductor channel opens for conductivity at positive gate voltages, with a threshold voltage of roughly $V_{th}$ = 43 V. Figure S4b shows low-temperature *I-V* characteristics acquired at different gate voltages. While the Cr/Au electrodes yield nearly ohmic *I-V* characteristics at room temperature (shown in Supplementary Note 5), the ohmic response is lost at $T$ = 7 K, where we obtain asymmetric, nonlinear *I-V* characteristics, indicating the presence of small but non negligible Schottky barriers at the Cr-ReS$_2$ interfaces.

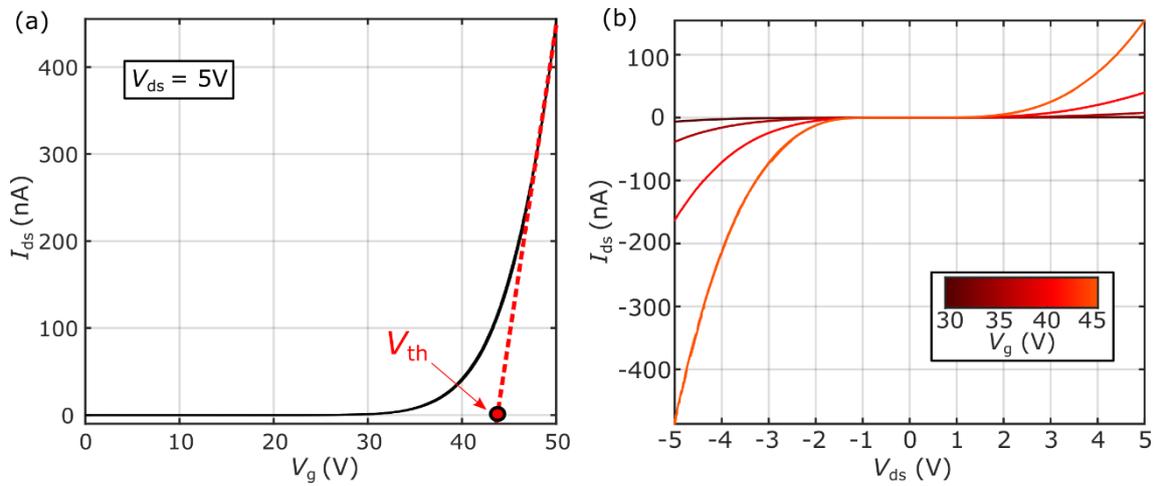

**Figure S4.** Electrical characterization of the device at T = 7 K. (a) Gate trace of the device (b) I-V curves at four different gate voltages.

## 5. Characteristic times of photoresponse in the device

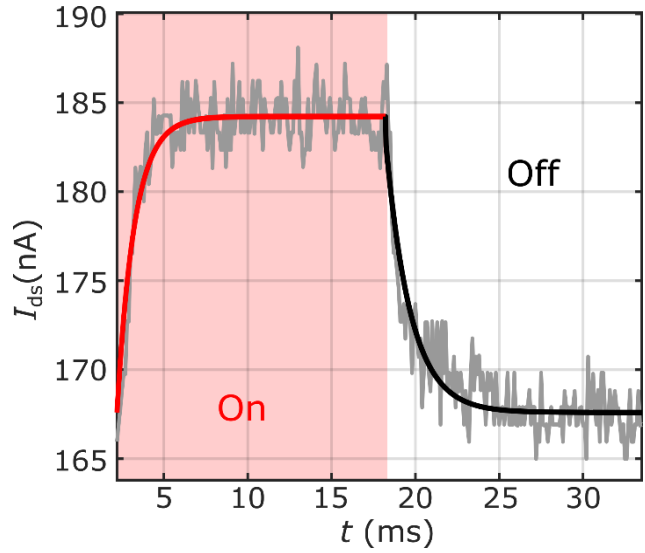

**Figure S5**. Time-dependent photoresponse of the device. Detail of a single on-off measurement of the device (gray line). Fitting of the on (red line) and off (black line) to an exponential increase or decrease of the photocurrent.

Figure S5 shows the time evolution of the source-drain current, $I_{ds}$ at $V_{ds}$ = 5 V and $V_g$ = 45 V for a on-off cycle of the illumination. The optoelectronic response after turning on the illumination can be well reproduced by a exponential model of the form

$$I_{ds}(t) = I_{off} + I_{ph,1}\left(1 - e^{-t/\tau_r}\right)$$

where $I_{off}$ is the drain-source current prior to illumination and $I_{ph,1}$, $\tau_r$ are used as fitting parameters, being $\tau_r$ the characteristic time of response. Similarly, the time evolution of $I_{ds}(t)$ immediately after turning of the light can be modeled as

$$I_{ds}(t) = I_{off} + I_{ph,2}e^{-t/\tau_d}$$

being $I_{ph,2}$ and $\tau_d$ the fitting parameters. We obtain the characteristic times of response of the device, $\tau_r$ = 1.05 ms and $\tau_d$ = 1.48 ms.

## 6. Room-temperature measurements

We characterize the optoelectronic response of the device at room temperature. Figure S6a shows the transfer curve of the device measured at $V_{ds}$ = 2 V. Due to leakage problems in the transitor we keep the gate voltage over – 35 V. We calculate the threshold voltage of the device fitting the linear part of the transfer curve (red dashed line). Figure S6b shows the I-V characteristics of the device at different gate voltages. We expose the device to light to measure the spectral dependence. Figure S6c shows the photocurrent spectra at different light polarization angles, from 0° to 180° in steps of 20°. Excitonic features broaden because of thermal energy, making it difficult to clearly distinguish the different optical transitions. Figure S6d shows the variation of the photocurrent as a function of the light polarization at a fixed energy of E=2 eV. Blue solid line shows the fitting of the experimental data to $I = I_0 + I_1 \cos^2(\theta + \theta_0)$. We observe a clear linear dichroic photoresponse of the device.

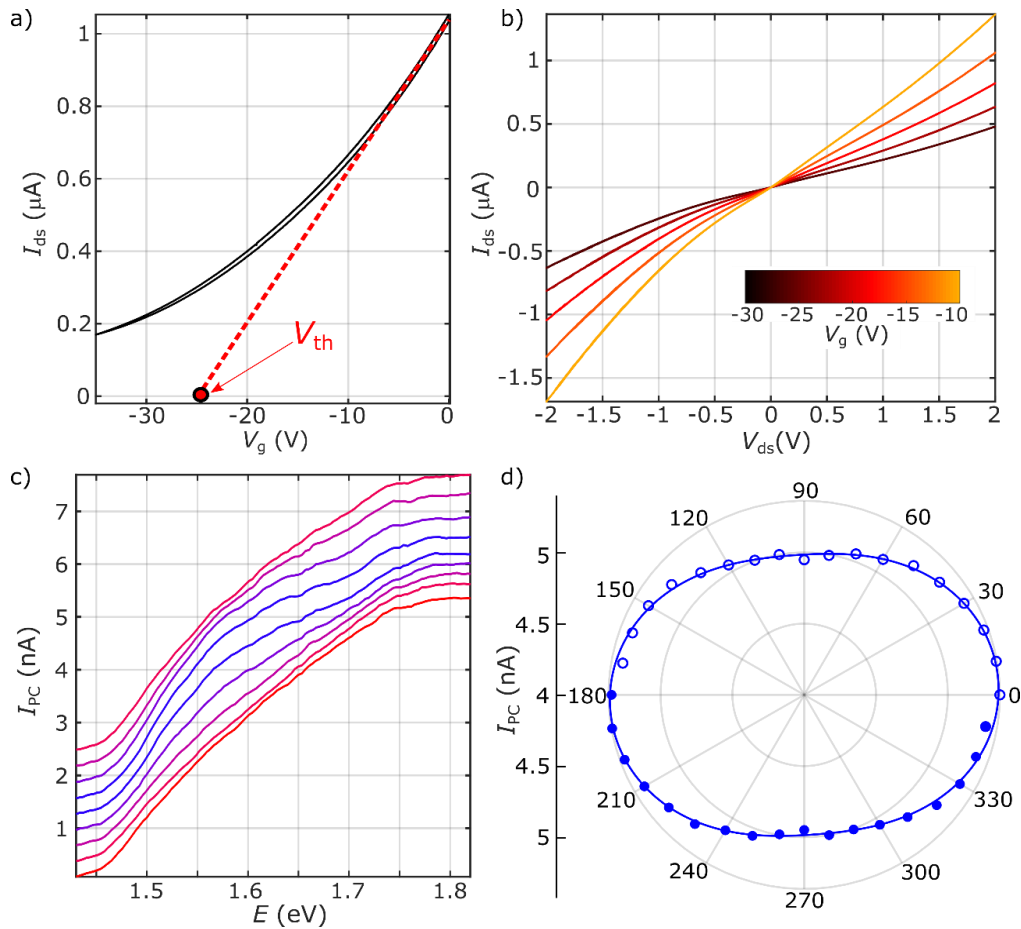

**Figure S6.** Room-temperature electrical characterization of the device. a) Gate transfer curve of the device. b) I-V characteristics measured at different gate voltages. c) Photocurrent spectra at different polarization angles from 0 to 180° in steps of 20°. d) Dichroism of the photocurrent measured at 1.8 eV.

## 7. Experimental results for AA stacked sample

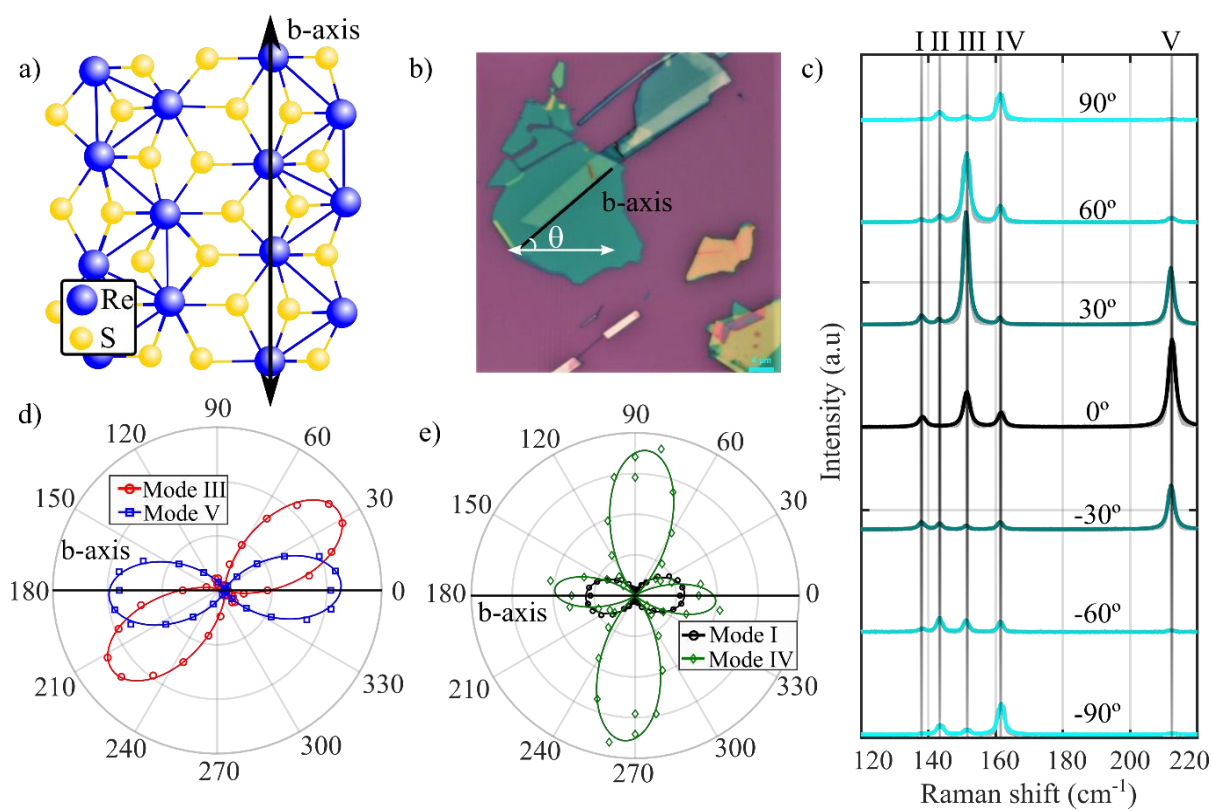

**Figure S7.** Polarization-resolved Raman spectroscopy for AA-stacked flake. a) Crystalline structure of ReS$_2$. b) Picture of the few-layer ReS$_2$ flake transferred on bottom-hBN. The definition of the angle between the light polarization (white arrow) and the *b*-axis (red line) is also shown, which will be used in the following measurements. c) Polarization-resolved Raman spectra as function of the sample orientation angle. The spectra taken every 10º between 0º and 180º are vertically offset. Mode I, II and IV are marked with dashed green lines. Mode I to V are labelled in the figure. d) Raman intensity of modes III (red circles) and V (blue squares) as a function of the polarization angle. e) Raman intensity of modes I (black circles) and IV (green diamonds) as a function of the polarization angle.

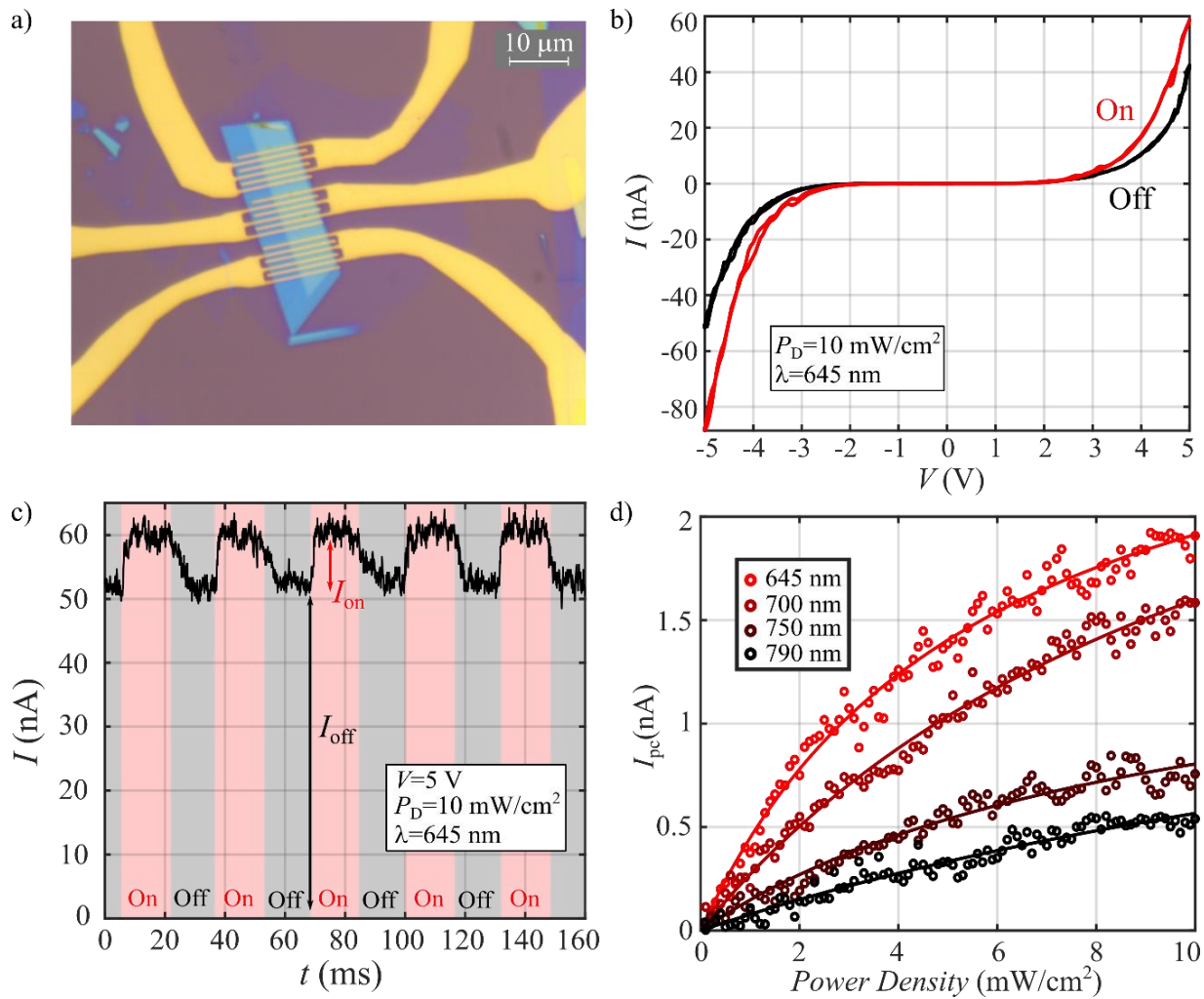

**Figure S8.** Optoelectronic characterization of the device. a) Optical image of the ReS$_2$ photodetector b) I-V curves with monochromatic illumination of 645 nm (red solid line) and dark (black solid line). c) Source-drain current of the photodetector at V$_{ds}$ = 1V. When the light excitation is turned on, the drain-source current increases by I$_{PC}$ = 1.5 nA. d) Power dependence of I$_{PC}$ at different wavelength excitation for V$_{ds}$ = 1V.

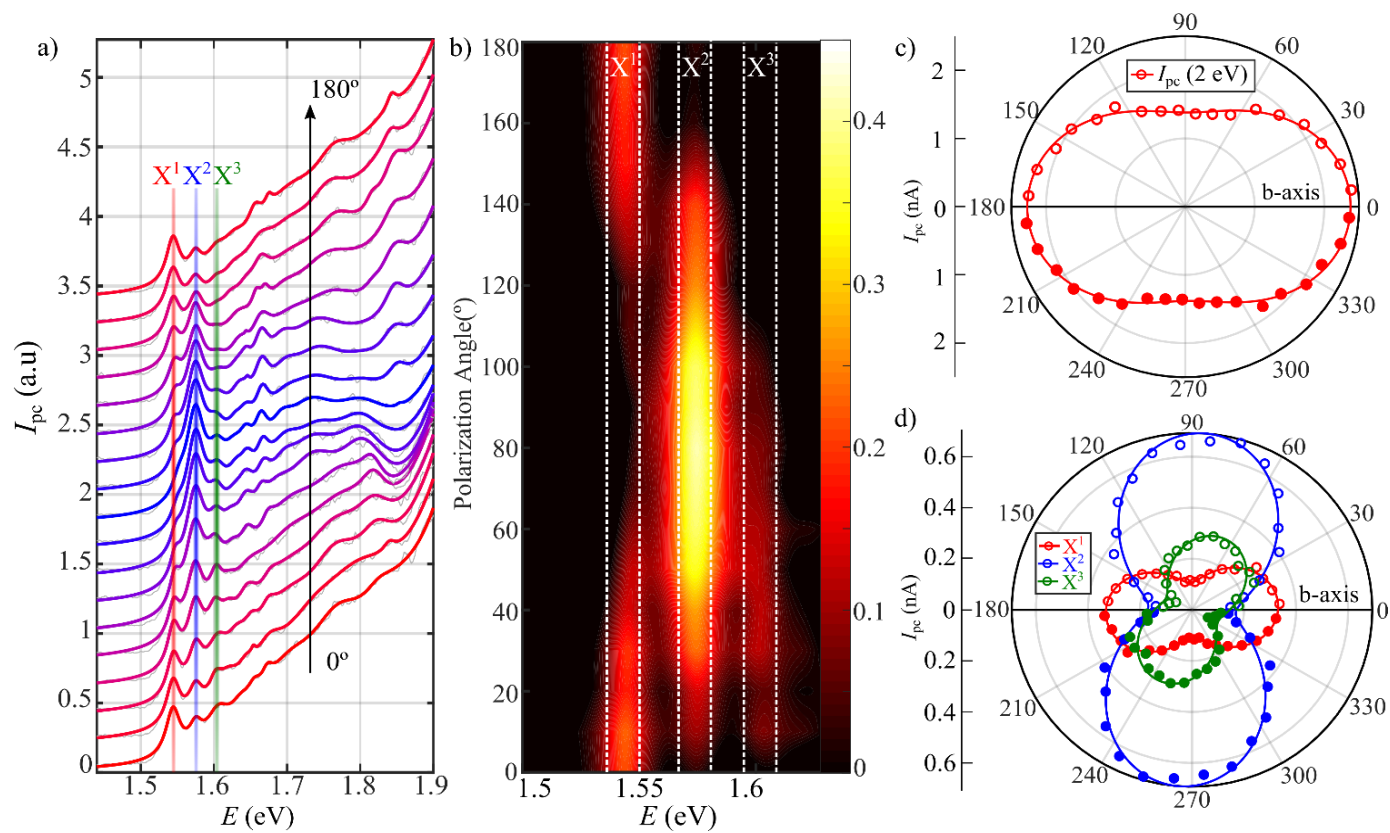

**Figure S9.** Polarization-resolved photocurrent spectroscopy. a) Polarization-resolved photocurrent spectra as function of the polarization of incident light. Red, blue and green lines mark the three neutral excitons. b) Color-map of the excitonic modulation c) Photocurrent as function of the angle of the polarization light with the *b*-axis at 2 eV. d) Amplitude of the exciton I, II and III as a function angle of the polarization light with the *b*-axis. Solid lines correspond to the fittings of the data.

# 8. Procedure for spectral acquisition

Figure S10a shows the photocurrent of the device as a function of the illumination power, at $V_{sd}$ = 5 V and $V_g$ = 45 V a modulation frequency of f = 31.81 Hz, and three different excitation energies. The photocurrent shows a sublinear power dependence, which can be well fitted by the power law $I_{PC} \propto P^\alpha$ obtaining roughly α~0.6. This trend suggests that the main mechanism of the generation of the photocurrent is the photogating effect. Figure S10b shows the spectral density of the light source used for acquisition of photocurrent spectra. Since the excitation power fluctuates with the wavelength, it is necessary to correct the acquired spectra accordingly. Figure S10c depicts the responsivity spectrum of the device with a light polarization angle of 0° respect the b-axis, before (black curve) and after (red curve) correcting for power fluctuations. The measured photocurrent is given by $I_{PC}(P_{lamp}) \propto P_{lamp}^\alpha$. Thus, the photocurrent at a given power $P$ can be obtained as $I_{PC}(P) = I_{PC}(P_{lamp})(P/P_{lamp})^\alpha$. The responsivity will be given by $R(P) = I_{PC}(P)/P = I_{PC}(P_{lamp})P^{\alpha-1}P_{lamp}^{-\alpha}$.

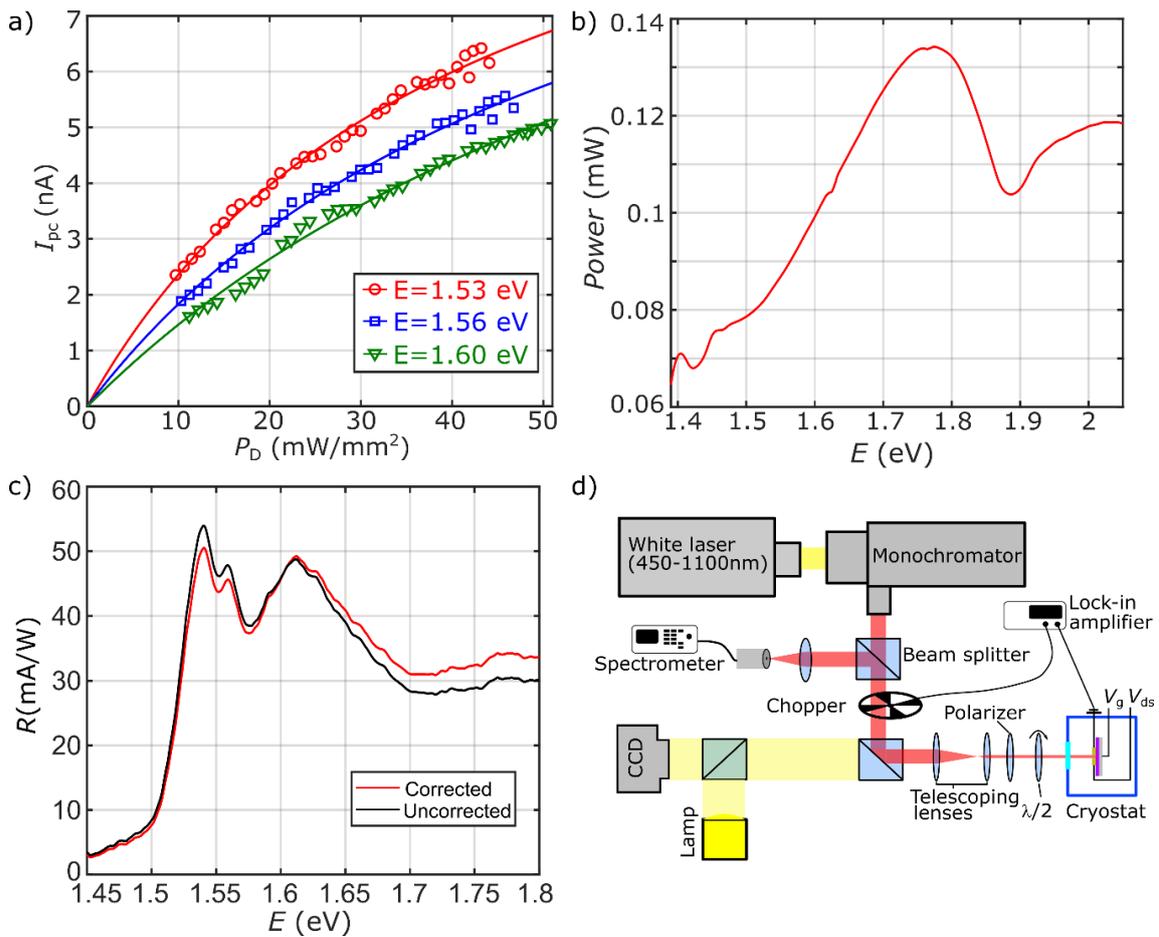

**Figure S10.** Procedure of spectral acquisition. a) Power density dependence of the photocurrent for different excitation energies. Solid lines correspond to fittings of the experimental data to $I_{PC} \propto P^\alpha$ b) Power spectrum of the light source. c) Corrected and uncorrected responsivity spectra of the device. d) Schematic of the low temperature photocurrent spectroscopy setup.

Figure S10d outlines the experimental setup. The sample is placed inside a closed cycle cryostat at T=7 K with an optical access and exposed to laser illumination. The light source is a supercontinuum (white) laser (SuperK Compact), and the excitation wavelength is selected using a monochromator (Oriel MS257 with 1200 lines/mm diffraction grid). This allows to scan the visible and NIR spectral range, roughly from 450 nm to 1000 nm. After the light source a beam splitter and a spectrometer allow us to measure the excitation wavelength. We select the polarization of the excitation using a linear polarizer and a half-waveplate placed in a rotating stage. In order to improve the signal-to-noise ratio of the optoelectronic measurements, the excitation signal is modulated by an optical chopper and the electrical response of the device is registered using a lock-in amplifier with the same modulation frequency. We visualize the sample using a lamp and a CCD camera.

## 9. First-principles simulations

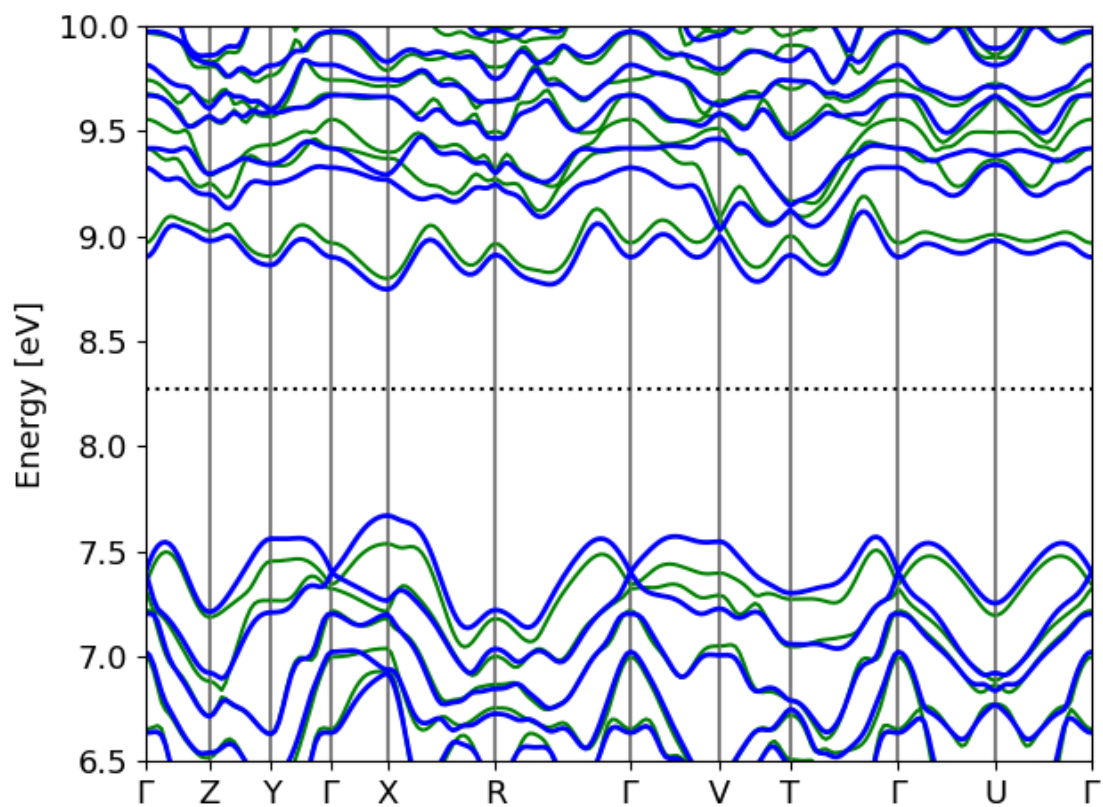

**Figure S11.** Band structure of bulk ReS$_2$ with (blue) and without (green) spin-orbit coupling for a path along the whole Brillouin zone, using the PBE exchange-correlation functional.